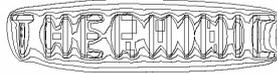

# A COUPLED THERMOREFLECTANCE THERMOGRAPHY EXPERIMENTAL SYSTEM AND ULTRA-FAST ADAPTIVE COMPUTATIONAL ENGINE FOR THE COMPLETE THERMAL CHARACTERIZATION OF THREE-DIMENSIONAL ELECTRONIC DEVICES: VALIDATION

*Peter E. Raad, Pavel L. Komarov, and Mihai G. Burzo*

Nanoscale Electro-Thermal Sciences Laboratory
Department of Mechanical Engineering
Southern Methodist University
Dallas, TX 75275-0337, U.S.A

## ABSTRACT

This work builds on the previous introduction [1] of a coupled experimental-computational system devised to fully characterize the thermal behavior of complex 3D submicron electronic devices. The new system replaces the laser-based surface temperature scanning approach with a CCD camera-based approach. As before, the thermo-reflectance thermography system is used to non-invasively measure with submicron resolution the 2D surface temperature field of an activated device. The measured temperature field is then used as input for an ultra-fast inverse computational solution to fully characterize the thermal behavior of the complex three-dimensional device. For the purposes of this investigation, basic micro-heater devices were built, activated, and measured. In order to quantitatively validate the coupled experimental-computational system, the system was used to extract geometric features of a known device, thus assessing the system's ability to combine measured experimental results and computations to fully characterize complex 3D electronic devices.

## 1. INTRODUCTION

Faster and more powerful devices mean hotter devices, which can lead to a decrease in performance and reliability. Thus, understanding and determining the thermal behavior of modern electronics has become a key issue in their design. As a result, there is a critical demand for methods that can be used to determine the temperature of features at the submicron level, particularly when most important features are physically inaccessible [2-5]. Computational approaches can provide insight into the internal thermal behavior of such complex devices, but can be limited by the inherent necessity of modeling the heat sources, which in the case of self-heating microelectronic devices, are the result of electrical fields whose exact shapes and locations are difficult to specify with reasonable certainty. Moreover, such devices can actually experience irreversible changes in thermo-physical properties and/or geometries that cannot be otherwise predicted from theory or monitored. Experimental approaches can also be helpful in determining thermal behavior, but require either physical access or a visual path to the region of interest. Contact methods, for example, present the difficulties of having to access features of interest with an external probe, or in the case of embedded features, fabricate a measuring probe into the device, and then having to isolate and exclude the influence of the probe itself. Non-contact methods, on the other hand, can provide surface temperature profiles, but in and of themselves cannot impart information on internal behavior. In other words, these methods provide a two-dimensional perspective on what otherwise is, in the case of stacked complex devices, an intricate three-dimensional thermal behavior.

We show herein that by combining an experimental method capable of mapping the surface temperature of a complex device with high spatial and temporal accuracy together with a computational engine capable of rapidly and accurately resolving the geometric and material complexities of a full three-dimensional microelectronic device, it becomes possible to use the independent information from the experimental measurements to mitigate the lack of knowledge in the source model parameters, which directly affect the usefulness of the computational results. This article builds on the previous introduction by the authors of a proof of concept of a coupled computational-experimental approach that uses a measured two-dimensional surface temperature mapping to help obtain a fully three-dimensional thermal characterization of an active micro-device [1].

## 2. METHODOLOGY

The overall approach combines computational and experimental methods previously developed by the authors. The transient two-dimensional surface temperature is measured by the use of the thermoreflectance thermography system [6], while the three-dimensional thermal behavior of multi-layered integrated circuits (with embedded features) is inferred by solving the inverse heat transfer problem with the self-adaptive ultra-fast numerical technique [7, 8]. To minimize the number of uncertainty sources that the inverse method must deal with, the thermo-physical properties of the various thin layers can be measured independently with the Transient Thermo-Reflectance technique [9, 10].

### 2.1. Thermo-Reflectance Thermography System

The experimental temperature mapping system is based on the thermoreflectance (TR) method, where the change in the surface temperature is measured by detecting the change in the reflectivity of the sample. The measurement methodology requires two steps. First, the thermoreflectance coefficient must be determined for each of the surface materials in the measurement area. Second, the changes in the surface reflectivity as a function of changes in temperature are measured over the area of interest, with submicron spatial resolution. The resulting reflectivity data is combined to obtain a temperature field over that area of interest. The thermoreflectance coefficient, $C_{TR}$, varies as a function of the material under test, material layering, and the wavelength of the probing light [11]. Thus, in order to maximize the signal to noise ratio, it is important to use a light source whose wavelength produces the maximum value of $C_{TR}$ for the exposed material layer.

A schematic of the TR thermography (TRTG) system is shown in Fig. 1. The probing light reflects from the heated surface back along the optical path to the sensitive element of a CCD camera (512×512 pixels). The intensity of the reflected light depends on the reflectivity (temperature) of the sample's surface. The frames containing the change in surface reflectivity induced by the temperature variations of the DUT are acquired and scaled according to the calibrated data. The details of the TRTG method are presented in a companion article.

The calibration approach consists of determining the relationship between the changes in reflectance and surface temperature. The change in reflectance is measured by a differential scheme involving two identical PDs in order to minimize the influence of fluctuations in the energy output of the probing laser. The sample temperature is controlled by a thermoelectric (TE) element and measured with a thermocouple. The calibration must be performed for each of the materials on the surface of each device where a temperature mapping is carried out.

### 2.2. Computational Engine

The numerical engine is capable of simulating the transient thermal behavior of active multi-layered devices whose dimensions vary over several orders of magnitude and where the thermophysical properties of the materials used may not be isotropic. The thermal modeling engine is used for solving the required heat transfer problem of the corresponding physical device. The measured surface temperature field (outlined white square in Fig. 2) is then used as input signature for an optimization scheme that varies control parameters (e.g., source power, length) until the RMS error between the computed solution and the input signature is minimized.

The novel approach begins by solving the corresponding steady-state problem by the use of a grid nesting technique. Since the physical dimensions of the various materials used in modeling high performance electronic devices vary greatly, a uniform mesh that resolves all of the details in three dimensions results in a prohibitively large computational grid. A common method for dealing with dimensional variation is to skew the mesh and concentrate more grid points in areas where higher resolutions are needed. The shortcoming of using a biased-mesh approach to resolve the geometry is that the problem geometry, and not the temperature gradients, will end up dictating the meshing. The meshing strategy used in the development of the numerical engine [7] was set on ensuring that it is (i) automatic and adaptive, (ii) independent of user expertise, and (iii) independent of materials (including air), geometry features, embedded vias, and heat source locations. The approach makes it possible to start with the full 3D geometry of a real device in all of its complexity, and then uses a physics-based automatic error predictor to focus the entire available computational power on only those regions that require further refinement in order to achieve the level of acceptable error prescribed by the analyst [8]. The power of the method is that it uses effective thermal properties that are consistent with the local grid spacing at the particular grid level in use. As a result, dealing with air, embedded vias, and ultra-thin multi-layered structures requires no special treatment.

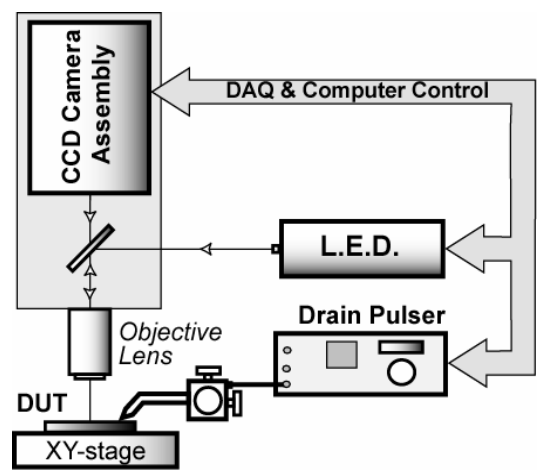

**Fig. 1  CCD-based TRTG scanning system**

### 2.3. Sample Characterization: Geometry and Thermal Properties

The fidelity of a computational solution depends directly on the accuracy of the material properties and geometric characteristics of the features making up the device of interest. Since it has been shown [12] that the thermal properties of thin-films vary from those of bulk materials, it is necessary for the numerical simulation to use the real values of the thermal conductivity for all of the materials making up the system under study. The previously described TTR measurement system [9] can be used to determine any unknown properties of thin-film materials and their interface resistances.

In order to carefully characterize the geometry of a device of interest, an ellipsometer was used to measure the thicknesses of transparent layers as well as to confirm the optical properties of surface metals. In addition, a profiler was used to measure the thicknesses of various opaque and transparent layers making up the device.

To prove the concept and validate the described method, we constructed basic reference aluminum micro-resistor devices buried in a layer of silicon dioxide, and investigated the accuracy with which the coupled experimental-computational approach is capable of determining key geometric features of a known device. In the micro-resistor device, shown schematically in Fig. 2, an aluminum (Al) strip heater is sandwiched between top and bottom oxide layers in the vertical direction, and between two Al activation pads in the horizontal direction. When activated at known electrical power levels, the known heat source will generate a three-dimensional temperature field throughout the device. Since the approach calls for measuring the temperature signature on the surface of the device, a gold (Au) layer was deposited over the anticipated area of interest, which includes the heat source and large surrounding regions. Gold was chosen since when used in conjunction with an LED light source at 485 nm, it will maximize the $C_{TR}$ value in the TRTG technique.

The simple construction of this micro-resistor device makes it possible to specify and measure all essential heat transfer problem parameters. Specifically, (i) the geometry of the different layers can be controlled in the fabrication process and later measured for confirmation; (ii) the oxide layers and Al strip provide a Joule heat source with known uniform power distribution; and (iii) the large pads make it possible to use a four-wire scheme to simultaneously activate the strip heater and measure its electrical power. For the purposes of this article, only a single device is reported on, which has a width of 14 μm and a length of 200 μm. All other pertinent geometric parameters are provided in Fig. 2.

### 3. RESULTS AND DISCUSSION

By using the experimental system and combining it with the numerical approach presented above it becomes possible to solve the inverse conduction problem associated with a complex, multi-layered, deep submicron electronic device in order to infer the thermal behavior of the embedded features that cannot be otherwise accessed. The inverse solution is obtained by varying key parameters that define the heat transfer problem under consideration. For the test micro-resistor in this work, these parameters include the size and location of the heater strip, its power and distribution, the thicknesses of the top and bottom layers, and their thermal properties. Obviously, a solution that comprehends all of these variables would be impractical. Furthermore, many of these parameters can be either directly measured (e.g., thermal properties, applied power, thicknesses of layers) and/or have smaller influences on the final temperature

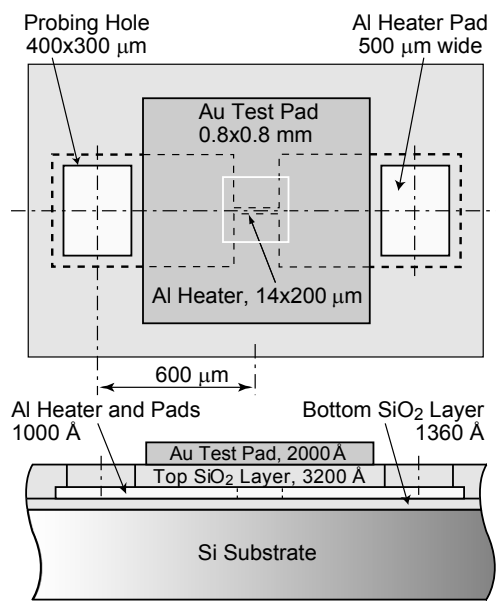

**Fig. 2 Geometry of test micro-resistor**

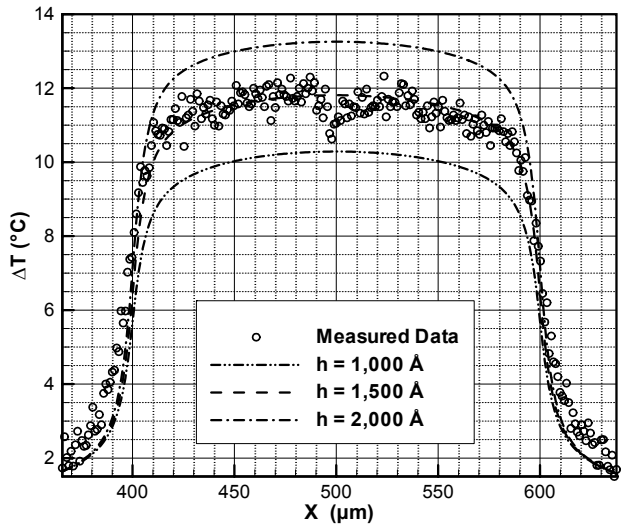

**Fig. 3** Influence of the bottom oxide thickness

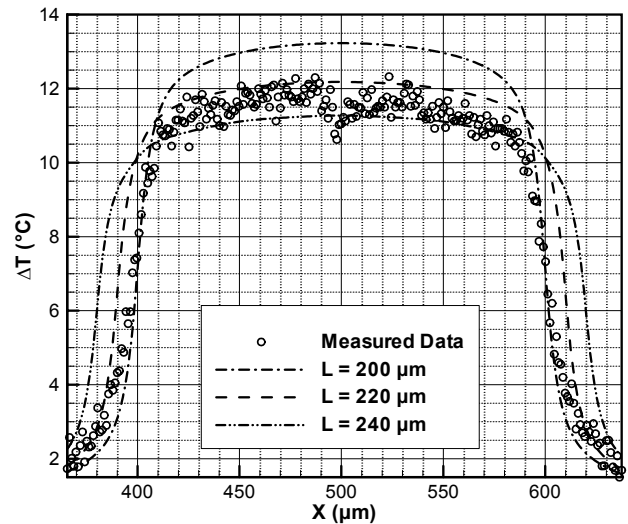

**Fig. 4** Influence of the heat source length

distribution in the device (e.g., layers of $SiO_2$ and Au above the heater).

Since heat flows primarily toward the substrate, the thickness of the bottom oxide layer is expected to strongly affect the temperature distribution because of the resistance that this layer presents. Therefore this parameter was chosen for its sensitivity on the final result. A second parameter was chosen to be the length of the heat source because (i) it cannot be directly measured and (ii) it is expected to be longer than the length of the heater strip itself because of the end effects at the junction between the aluminum heater strip and its activation pads. In fact, one would expect that the end effects would extend beyond the strip by a distance that scales with the heater's width.

Figure 3 compares the experimental data along the mid-plane of the heater strip to the corresponding numerically computed temperature distributions for different values of the bottom oxide thickness and for a fixed heat source length of 200μm. As expected, thicker layers of bottom oxide result in higher strip temperatures. It appears that the numerical temperature distribution for the thickness of 1,500Å fits nicely the measured data in the center of the heater, X∈[400,600], but misses inside the pad areas. The discrepancy is due to the end effect previously mentioned.

To examine the end effects, the bottom oxide thickness was held fixed at 2000Å, and the heat source length was varied as depicted by the results in Fig. 4. Indeed, extending the heat source length beyond the ends of the strip heater pulls the temperature curves outwardly toward the experimental data at the lower temperatures. However, the agreement between the numerical and experimental distributions becomes worse toward the center of the domain. It should be pointed out that simply extending the rectangular heat source, as done here, only captures the spherical nature of the heat distribution at the pad-heater junction to a first order approximation. In order to more precisely simulate the end effects, one would have to introduce a more sophisticated heat power distribution model in this region.

Nevertheless, taken together, the results of Figs. 3 and 4 point clearly to the importance of both physical parameters and hence the need to optimize over both of them simultaneously. The optimization method used in this work is a variant of the "steepest descent" method [13]. The search begins with a numerical solution of the heat transfer problem at nominal values of the two parameters being considered, i.e., bottom oxide thickness (h) and heat source length (L). This numerical solution is compared with the measured signature field (outlined white square in Fig. 2) on the gold pad at every common location to compute an RMS error. Then, additional trial solutions and associated errors are obtained at neighboring pair values of h and L in the two-dimensional parameter space. Evaluation of the errors at the nominal and eight neighboring pairs provides the direction for modifying h and L in order to decrease the error. This process continues until the changes in both h and L are acceptably small.

For the specific device under consideration, Fig. 5 shows the path taken in the (h, L) parameter space to converge onto the final values of h and L which yield a numerical solution that matches the measured surface signature to within the minimum RMS error. Of particular interest here is that the process of reaching the final result required 57 solutions of the full three-dimensional heat transfer problem, each of which was converged to less than 1% numerical error at the specified h and L values. It is obvious that such an approach would be impractical with traditional numerical solvers. The self-adaptive, ultra-fast computational engine used here required approximately 20 minutes to solve this particular problem on a 3.4 GHz Pentium(R) 4 desktop PC.

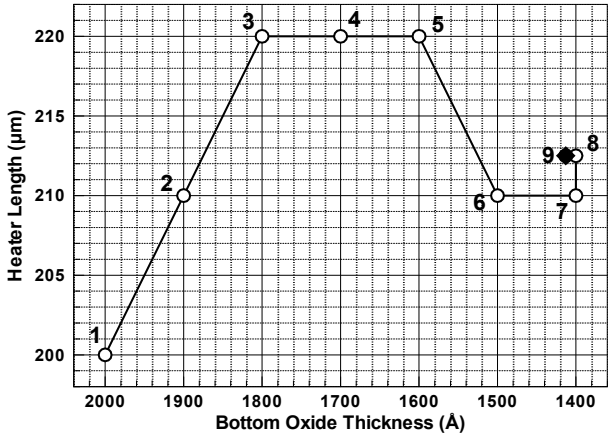

**Fig. 5** Convergence of optimization path toward final solution at h = 1,413 Å and L = 212.5 μm

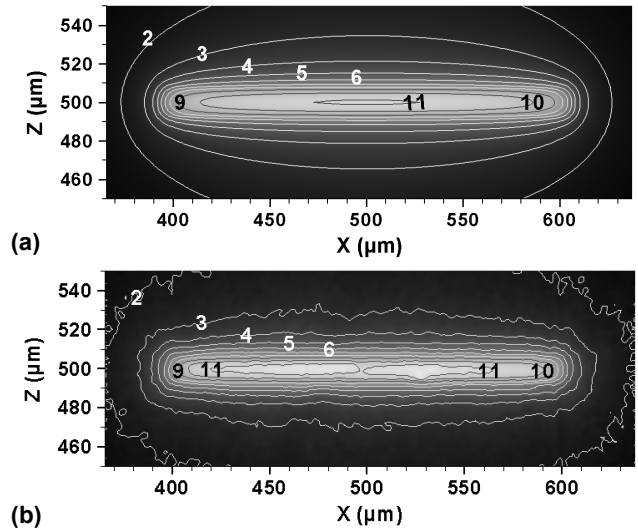

**Fig. 6** Contours of surface temperature: (a) optimal numerical solution and (b) experimental signature

To compare the experimental and numerical surface temperature fields, a surface slice is extracted from the full 3D solution that corresponds to the size, location, and resolution of the experimental area. Figure 6(a) shows the surface temperature slice at stage 9, which represents the "optimal" solution that is closest to the experimental signature shown in Fig. 6(b). While the agreement is very favorable, it is clear that the end effects give the experimental contours a more "rectangular" shape on the heater edges. As discussed above, a more sophisticated model of the power distribution at the pad-heater junction would be required to further improve the agreement, which is beyond the scope of this validation article.

The ultimate benefit of the coupled experimental-numerical system is that it takes an inherently 2D experimental approach and provides a full 3D thermal characterization of the complete device. The optimization-based coupling ensures that the 3D solution is consistent with the 2D experimental signature within the chosen heat transfer model as defined by the combination of known and unknown (i.e., optimizable) parameters. Figure 7 provides an example of the optimal 3D numerical solution. For clarity, only two slices are shown and the spatial domain is restricted to the region surrounding the heater, even though the computational domain is 1000×500×1000 μm. The horizontal slice is on the surface of the device and the vertical slice cuts across the mid-plane of the heater.

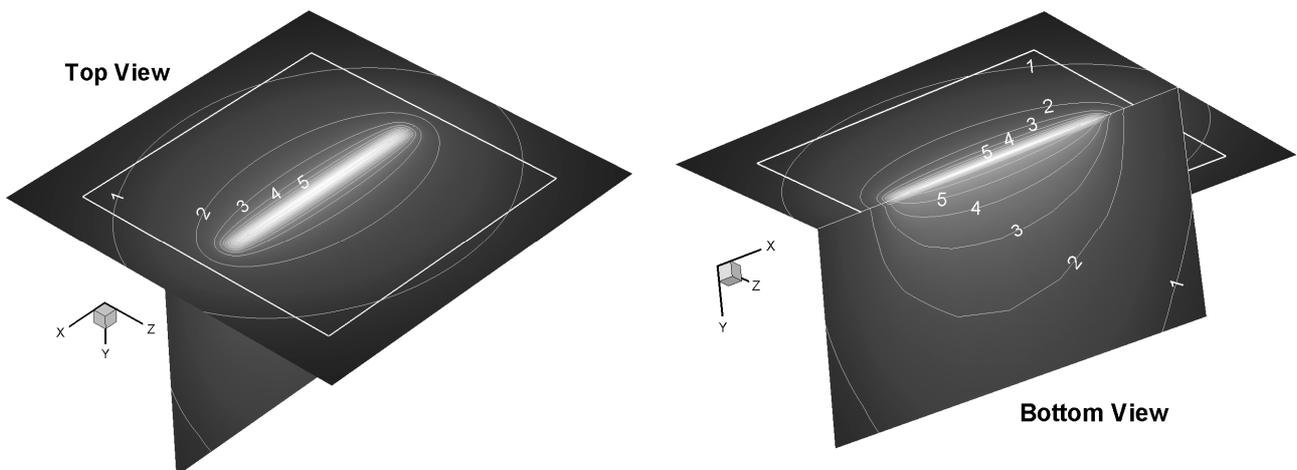

**Fig. 7** Contours of 3D temperature field from optimal numerical solution (stage 9 in Fig. 5). The solid white square delineates the 272 μm square measurement surface area, also shown on Fig. 1.

## 4. CONCLUSIONS

This work builds on the previous introduction [1] of a coupled experimental-computational system that makes it possible to extend an experimentally-obtained 2D surface field to a full 3D characterization of the thermal behavior of complex submicron electronic devices. The new system improves on the previous one by replacing the laser-based, point-by-point, surface temperature scanning approach with a CCD camera-based approach. The aim of this work was to validate the coupled system by applying it to a simple test micro-resistor device with known geometric, material, and thermal characteristics. The investigation focused on the two parameters that most influence the heat transfer solution within the device, which are the thickness of the bottom oxide (h) and the length of the heat source (L). The obtained optimal value of h was found to be within 4% of the physical thickness as measured with an ellipsometer during the fabrication process. The optimal heat source length was found to extend into the side pads by a distance that is approximately equal to half the heater's width, which is qualitatively consistent with the physics of Joule heating in the sudden expansion areas connecting the strip and the pads. The results herein have provided the desired validation for the coupled experimental-computational system as well as demonstrating the power of the method in providing full 3D temperature field that agrees very well with the measured 2D surface temperature signature.

For existing devices, the highly resolved and accurate 3D temperature field would provide the ability to detect hot spots, diagnose performance, and assess reliability. In the design and manufacturing of new devices, this new tool has the potential to provide a rapid approach for analyzing the thermal behavior of complex stacked structures, to identify regions of excessive heat densities, and ultimately to contribute to improved thermal designs, better device reliability, and shorter design cycle time.

## 5. ACKNOWLEDGEMENTS

We wish to thank Mr. Jay Kirk of the EE Department at SMU for his help in fabricating the devices for this work.